# Demonstration of an 8D Modulation Format with Reduced Inter-Channel Nonlinearities in a Polarization Multiplexed Coherent System

A. D. Shiner,[*] M. Reimer, A. Borowiec, S. Oveis Gharan, J. Gaudette, P. Mehta, D. Charlton, K. Roberts and M. O'Sullivan

*Ciena Corporation, 3500 Carling Ave., Ottawa, Ontario, Canada*
[*]*ashiner@ciena.com*

**Abstract:** We demonstrate a polarization-managed 8-dimensional modulation format that is time domain coded to reduce inter-channel nonlinearity. Simulation results show a 2.33 dB improvement in maximum net system margin (NSM) relative to polarization multiplexed (PM)-BPSK, and a 1.0 dB improvement relative to time interleaved return to zero (RZ)-PM-BPSK, for a five channel fill propagating on 20x80 km spans of 90% compensated ELEAF. In contrast to the other modulations considered, the new 8-dimentional (8D) format has negligible sensitivity to the polarization states of the neighboring channels. Laboratory results from High-density WDM (HD-WDM) propagation experiments on a 5000 km dispersion-managed link show a 1 dB improvement in net system margin relative to PM-BPSK.

---

## 1. Introduction

The high cost of capacity for transoceanic optical transmission has spurred development of new terminal equipment solutions for best effect on existing dispersion managed links. For ultra-long haul applications, coherent detection with PM-BPSK [1] and time interleaved return to zero BPSK [2, 3] are state of the art solutions. In many cases capacity is limited by power and polarization dependent Kerr nonlinear (NL) interactions within and between channels. Eventually, NL interference will be partially compensated with data-dependent compensation fields applied at the transmitter or receiver [4, 5]. At present, the complexity of such solutions limits the commercial applicability of this approach. Current implementable solutions for mitigating NL interference include the use of linear pre-dispersion applied at the transmitter [6, 7] and rapid alternation of the transmit polarization state [8] to minimize cross-polarization modulation (XPolM). The linear (and NL) performance of a channel is also strongly influenced by the constellation design. In the linear propagation regime, noise tolerance is improved through coding techniques which minimize the symbol error rate by exploiting the dimensionality of the signaling space [1]. The development of high spectral efficiency formats remains an active area of research [9], however most cases focus on minimizing sensitivity to noise rather than reducing its generation all together.

In this paper, we propose and demonstrate a novel method for time-domain coding that substantially reduces inter-channel NL interference. Constellation symbols span the 8-dimensional space defined by two adjacent signaling intervals (time slots), and the four dimensions of the complex optical field within each time slot. The encoding of the cross-polarized, 'X' modulation format maintains the same spectral efficiency as PM-BPSK, while increasing the minimum Euclidean distance by encoding four bits across two time slots. The format is power balanced with polarization symmetry that reduces polarization scattering effects [10]. The constellation alphabet is chosen so that within each symbol the polarization Stokes vectors for the two time slots are equal and opposite. Thus the degree of polarization

(DOP) of the symbol is zero. Depending on the application, this solution is shown to provide up to a 2.3 dB increase in net system margin at an un-coded BER of 3.5% relative to PM-BPSK at the same spectral efficiency. We report a measured 1.0 dB improvement for an HD-WDM configuration, with 37.5 GHz channel spacing, propagating on a 5000 km dispersion managed link.

## 2. Modulation Format Design and Nonlinear Propagation Characteristics

We compare modulation formats in terms of net system margin (NSM), which is defined as the difference between the channel's optical signal-to-noise ratio (OSNR) and the OSNR required to maintain a specified un-coded bit error rate. A simple Gaussian noise (GN) model [11] is used to estimate the performance difference between two modulation formats at the powers that maximize their respective NSMs. The difference in maximum NSM, $\Delta NSM_{max}^{dB}$, is

$$\Delta NSM_{max}^{dB} = -\frac{3}{2}\Delta SNR_{b2b}^{dB} + \frac{1}{2}\Delta SNR_{NL}^{dB} \qquad (1.1)$$

where $\Delta SNR_{b2b}^{dB}$ is the difference in back-to-back (B2B) required SNR and $\Delta SNR_{NL}^{dB}$ is the difference in NL SNR following propagation at an arbitrary reference power. For equivalent implementation impairments, the first of these terms is determined by the minimum Euclidean distances of the modulations, and the second by their nonlinear noise characteristics. We will show here that the choice of coding can have a dramatic impact on NL noise generated during propagation.

Recently, a 4-ary frequency and polarization switched QPSK format (4FPS-QPSK) was demonstrated [12] where symbols were encoded across an 8D space comprising two optical subcarriers. The 4-bit per symbol solution was compared with dual carrier polarization multiplexed QPSK (8 bits per symbol) and achieved an impressive $\Delta SNR_{b2b}^{dB} \approx 5$ dB (3 dB of which is due to the different spectral efficiencies). From the first term in (1) we would expect a ~ 7.5 dB improvement in maximum NSM if both formats engender the same degree of NL interference during propagation. However, a reach increase of 84% ($\Delta SNR_{max}^{dB} \approx 2.6$ dB) was measured, which suggests that the NL propagation penalty of 4FPS-QPSK overwhelmed its sizable back-to-back advantage. The 4FPS-QPSK performance shortfall was attributed to difficulty with phase tracking as well as possible difference in NL propagation performance.

The X-constellation is a particular variant of an 8D bi-orthogonal constellation [12]. We start with a standard bi-orthogonal constellation defined as all coordinate permutations of [±1,0,0,0,0,0,0,0], where the first four coordinates are sent in time slot (A) and the second four are sent in slot (B). Through 8D Euclidean distance decoding this format achieves a 0.53 dB coding gain at an un-coded bit error rate (BER) of 3.5% as compared to four dimensional decoding of PM-BPSK which shares the same 2 bit per time slot spectral efficiency.

Noting that XPM and SPM processes are governed by the time dependent change in the power of the interfering channel, and XPolM additionally by its change of polarization state, it is clear that the switching characteristic of the bi-orthogonal constellation as defined above is disadvantageous for NL propagation. To improve the format's propagation characteristics we apply a series of 8D rotations, taking advantage of the fact that the Euclidean distance between constellation points is invariant under orthogonal transformations. We first rotate the constellation to ensure constant symbol energy for both time slots. To address XPolM, we identify a rotation where the two time slots that define each symbol have equal and opposite polarization Stokes vectors, guaranteeing that the symbol's DOP is zero. This last rotation removes any sensitivity to the relative polarization states between neighboring channels. The resulting constellation alphabet is summarized in Table 1. While the energy and polarization symmetry properties are shared with other space-time block codes, for example, the

Alamouti-QPSK format [13], our X-constellation provides an additional 0.53 dB coding gain over existing methods.

**Table 1.** Jones vectors for each of the consecutive time slots (A and B) that define the X-constellation symbols. The Stokes vectors for each pair of time slots are reported as $S_A$ and $S_B$, respectively. Note that the cumulative Stokes vector $S_A+S_B$ is zero for all symbols.

| Binary Value | | 0000 | 0001 | 0010 | 0011 | 0100 | 0101 | 0110 | 0111 | 1000 | 1001 | 1010 | 1011 | 1100 | 1101 | 1110 | 1111 |
|---|---|---|---|---|---|---|---|---|---|---|---|---|---|---|---|---|---|
| Slot A | x-pol | -1-i | -1-i | -1-i | -1-i | -1+i | -1+i | -1+i | -1+i | 1-i | 1-i | 1-i | 1-i | 1+i | 1+i | 1+i | 1+i |
| | y-pol | 1+i | -1-i | 1-i | -1-i | -1-i | 1-i | -1-i | 1+i | -1-i | 1-i | -1+i | 1+i | 1+i | -1+i | 1-i | -1-i |
| | $S_A=(S_1,S_2,S_3)$ | (0,-4,0) | (0,0,-4) | (0,0,4) | (0,4,0) | (0,0,4) | (0,-4,0) | (0,4,0) | (0,0,-4) | (0,0,-4) | (0,4,0) | (0,-4,0) | (0,0,4) | (0,4,0) | (0,0,4) | (0,0,-4) | (0,-4,0) |
| Slot B | x-pol | 1+i | -1-i | 1-i | -1-i | 1+i | -1+i | 1-i | -1-i | 1+i | -1+i | 1-i | -1-i | 1+i | -1+i | 1-i | -1-i |
| | y-pol | 1+i | -1-i | -1-i | 1+i | 1-i | -1-i | -1+i | 1-i | -1-i | 1-i | 1-i | -1+i | -1-i | 1+i | 1+i | -1-i |
| | $S_B=(S_1,S_2,S_3)$ | (0,4,0) | (0,0,4) | (0,0,-4) | (0,-4,0) | (0,0,-4) | (0,4,0) | (0,-4,0) | (0,0,4) | (0,0,4) | (0,-4,0) | (0,4,0) | (0,0,-4) | (0,-4,0) | (0,0,-4) | (0,0,4) | (0,4,0) |
| DOP | $\|S_A+S_B\|$ | 0 | 0 | 0 | 0 | 0 | 0 | 0 | 0 | 0 | 0 | 0 | 0 | 0 | 0 | 0 | 0 |

**2. Simulation Results:**

Simulation results comparing the NSM of the X-constellation to that of PM-BPSK and time interleaved RZ-PM-BPSK are shown in Figure 1. For these simulations we have adapted an extended reach RZ-PM-BPSK version of the RZ modulation described in [3], that is similar to the solution illustrated in [2]. All modulations were assessed with optimum linear pre-dispersion applied to each channel. We propagate a five channel fill with 50 GHz channel spacing over 20 x 80 km spans of 90% optically dispersion compensated ELEAF. To facilitate comparison with RZ-PM-BPSK, we set the symbol rate for all formats to 28 GSymb/s and optically filter each channel at the transmitter with a 42 GHz (FWHM) 4th order super-Gaussian filter. Following propagation, we noise load at the receiver and subsequently apply a filter optimally matched to the transmitter impulse response. A 31 tap MIMO filter is used to suppress residual inter-symbol interference prior to decoding. We note that optical filtering at the transmitter limits the power and polarization balanced properties of the RZ-PM-BPSK format because the RZ channel spectrum extends past 42 GHz. Results presented in Figure 1 correspond to the best and worst case polarization alignments for each format. For this simulated test case, X-constellation improves the NSM by 2.33 dB and 1.0 dB relative to the best-case polarization alignments of PM-BPSK and RZ-PM-BPSK, respectively, at an un-coded BER of 3.5%. As shown in Figure 1, the advantage of the X-constellation is considerably greater when budgeting for the worst case polarization alignment.

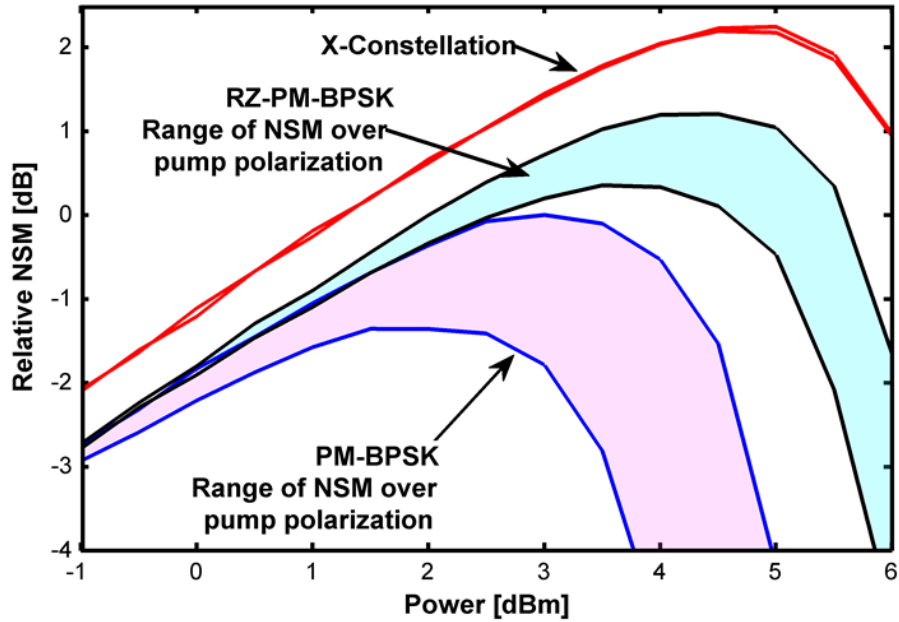

**Figure 1.** Simulated NSM, relative to the maximum NSM of PM-BPSK, for 5 channels with 50 GHz spacing propagated over 20x80 km spans of 90% optically dispersion compensated ELEAF.

**2. Laboratory Results:**

The NSM measurements for X-constellation and PM-BPSK were performed on the 5000 km dispersion and dispersion-slope managed link shown in Figure 2.(a), consisting of 76 spans of Corning Vascade® L1000, S1000 and EX2000 fiber with a residual end-of-link dispersion of -1500 ps/nm. As shown in Figure 2.(b), our test channel at 1550.1 nm was located in the center of 9 HD-WDM channels with 37.5 GHz spacing, in which each channel was modulated with an independent, commercially available transmitter (Ciena WaveLogic 3) operating at 35 GSymb/s. Optimum linear pre-dispersion was applied electronically to each HD-WDM channel. The remaining 51 channels, with 50 GHz spacing, were bulk modulated by a single modified transmitter followed by an optical channel de-correlator and eight continuous wave

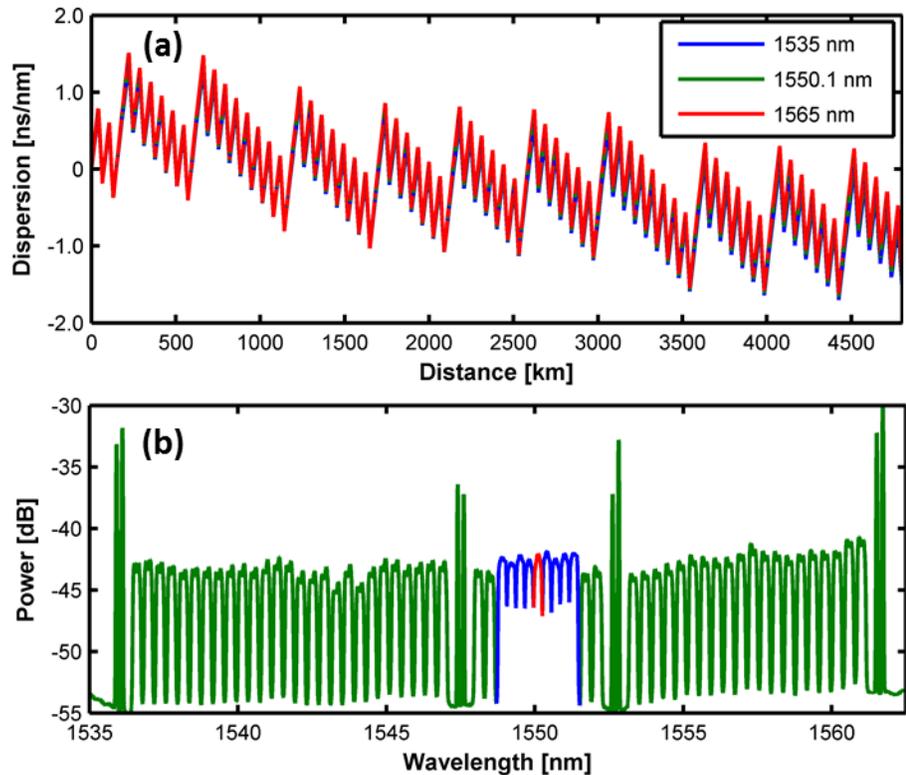

**Figure 2.** (**a**) The 5000km testbed dispersion map and (**b**) WDM channel configuration for the 5000 km fiber link. The 9 HD-WDM (37.5 GHz spaced) channels are illustrated in blue.

(CW) polarization scrambled lasers were used for power control. At the receiver, the test channel was optically noise loaded, detected with a standard coherent receiver and processed offline. Figure 3 shows the resulting NSM for the X-constellation and PM-BPSK at an uncoded BER of 3.5%. At low launch powers we observe a ~ 0.5 dB improvement in NSM associated with the Euclidean distance gain of the X-constellation. Furthermore, the X-constellation improved the maximum NSM by 1.0 dB relative to PM-BPSK after 5000 km of propagation at a spectral efficiency 33% higher than that afforded by conventional 50 GHz channel separations.

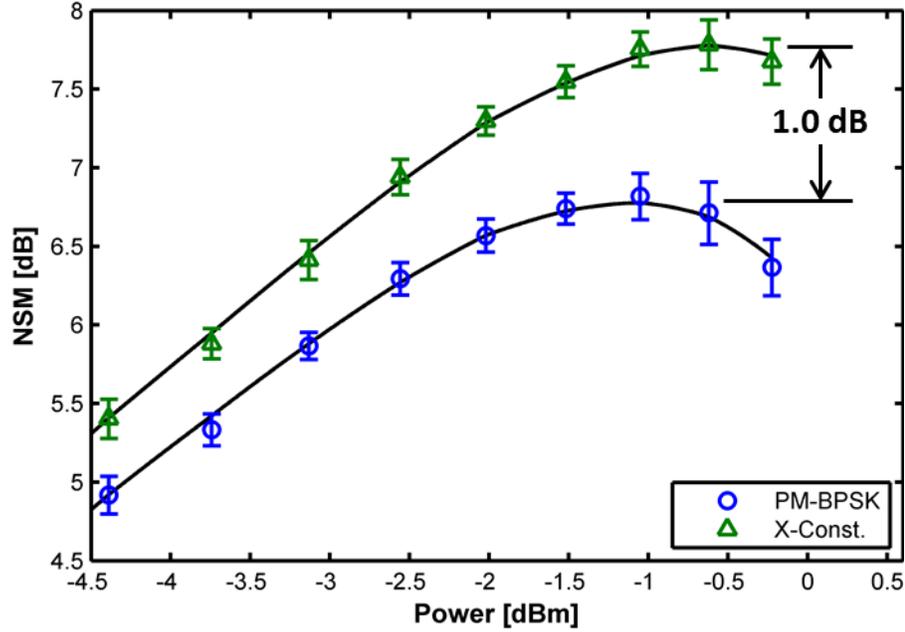

**Figure 3.** The measured NSM for PM-BPSK and X-constellation following 5000 km of HD-WDM propagation.

## 4. Conclusion

Whereas NL pre-compensation cancels the NL field that is generated during propagation, we have shown that coding can be used to reduce the generation of NL noise, while simultaneously increasing Euclidean distance, without sacrificing spectral efficiency. The implementation complexity of our solution is similar to that of PM-BPSK with the additional requirement that pairs of time slots must be encoded and decoded together. Experimentally, we realized a 1.0 dB improvement in NSM at 3.5% un-coded BER after 5000 km of HD-WDM propagation when comparing the X-constellation to PM-BPSK under identical propagation conditions and data rates. Simulation results show that the X-constellation outperforms PM-BPSK by 2.33 dB and RZ-PM-BPSK by 1.0 dB with greatly reduced sensitivity to the polarization states of neighboring channels. It has not escaped our notice that the method of encoding symbols across adjacent time slots and identifying transformations that minimize power variations and polarization asymmetry can be extended to higher spectral efficiency modulation formats.